\documentclass[3p,times]{elsarticle}

\usepackage{ecrc}

\volume{00}

\firstpage{1}

\journalname{Nuclear Physics A}

\runauth{}

\jid{nupha}

\usepackage{amssymb}

\usepackage[figuresright]{rotating}

\begin{document}

\begin{frontmatter}



\dochead{}

\title{Performance and First Physics Results of the ALICE Muon Spectrometer}

\author{Debasish Das (for the ALICE Collaboration)}

\address{Saha Institute of Nuclear Physics, 1/AF, Bidhan Nagar,
Kolkata 700064, India.\\
debasish.das@saha.ac.in , 
debasish.das@cern.ch
}

\begin{abstract}
A precise measurement of the heavy-flavor production cross-sections in pp collisions is an essential baseline for 
the heavy-ion program. In addition it is a crucial test of pQCD models in the new energy regime at LHC.
ALICE measures the muons from the decay of charmonium resonances and from the semileptonic decay of heavy-flavored 
hadrons in its forward (-4.0 $<$ $\eta$ $<$ -2.5) Muon Spectrometer. We discuss  the status of the detector 
and present results of data taken in pp collisions at $\sqrt{s}$=7 TeV.

\end{abstract}


\end{frontmatter}


\section{Introduction}
\label{}

A Large Ion Collider Experiment (ALICE)~\cite{Fabjan:2011jb} is a general-purpose heavy-ion experiment 
which has been designed to study the physics of the strongly interacting matter in 
nucleus-nucleus collisions at the Large Hadron Collider (LHC).
The ALICE experiment is the only experiment at the LHC devoted to heavy-ion physics to study the nature of the quark matter 
under the conditions of extreme temperature ($\geq$ 0.3 GeV) and high energy density ($>$ 10 GeV/fm$^{3}$).

The complicated structure of nuclear matter at low temperatures, where it is composed of a multitude of hadronic particles, 
baryons and mesons, is expected to give way at high temperatures to a plasma of weakly interacting quarks and gluons, 
the Quark-Gluon Plasma (QGP). A thermalized system where the properties of the system are governed by the quark and 
gluon degrees of freedom is called the QGP~\cite{Shuryak:1978ij}. Heavy quarks (charm and beauty) are produced in the first stages of the relativistic collisions and then they coexist with the surrounding medium due to 
their long life-time~\cite{Braaten:1991we}. Transverse momentum ($p_{T}$) and rapidity ($y$) distributions and 
quarkonia~\cite{Matsui:1986dk,physreptvogt} production rates are the significant probes which 
will allow for probing the properties of the medium. 

The LHC will allow us to probe the parton distribution functions of the nucleon and in the case of pA and AA collisions,
also their modifications in the nucleus, at very low values of momentum fraction (Bjorken $x$). Due to its lower mass, 
charm allows us to probe in lower $x$ than beauty. The capabilities to measure charm and beauty particles in the 
forward rapidity region ($|y|\simeq4$) using the Muon Spectrometer~\cite{tdr} gives access to the 
regime of $x\sim10^{-6}$. The ALICE Muon Spectrometer 
physics program is focussed on the measurement of heavy-flavor production in pp, pA and AA collisions at LHC energies.

Heavy-quark production measurements in pp collisions at LHC energies provide the necessary baseline for 
the study of medium effects in AA collisions along 
with the test of perturbative QCD (pQCD) in an unexplored energy regime much higher than the RHIC energies~\cite{Fabjan:2011jb}. 
The strong interaction with the high density QCD matter results in an energy-loss~\cite{Matsui:1986dk} whose 
effects were already seen in RHIC results~\cite{Braaten:1991we}, but further analyses are required to understand the details of 
underlying mechanism. Also the production mechanisms of the various theoretical models~\cite{physreptvogt,tevatron} needs 
to be tested and the relevant observables (from the experience in $p\overline{p}$ collisions at 
Tevatron energies~\cite{tevatron}) are quarkonia cross-sections and $p_{T}$ distributions.

The data taking in the ALICE experiment started from 2009 with 
pp collisions at the LHC at $\sqrt{s}$ = 900 GeV and from March 2010 at $\sqrt{s}$ = 7 TeV. 
Data for Pb-Pb collisions at $\sqrt{s_{{NN}}}$ = 2.76 TeV were successfully taken from November 
7 to December 6, 2010. 

\section{Heavy-flavor detection with the ALICE Muon Spectrometer and pp data taking in 2010}
\label{}

In the framework of the ALICE physics program~\cite{Fabjan:2011jb}, the goal of the Muon spectrometer of
ALICE is the study of quarkonia production, open heavy flavor production and vector meson properties 
via the muonic decay channel. For heavy-ion collisions the dependence with the collision centrality and with the 
reaction plane will also be studied. The spectrometer acceptance covers the forward 
(LHC-clockwise stream with respect to the interaction point) pseudo-rapidity domain 
of -4.0 $<$ $\eta$ $<$ -2.5 (small angles between $2^{0}$ and $9^{0}$ relative to the 
beam direction). The resonances can be detected down to zero transverse momentum.

The Muon spectrometer is comprised of absorbers to filter the background, a 
10 plane tracking chamber system before, inside and after the magnet, a muon filter (iron wall) 
and a set of trigger chambers. Figure. 1 shows the layout of the ALICE Muon Spectrometer. 
Outside the L3 magnet there is a large warm dipole magnet 
with a 3 T$\cdot$m field integral. Absorbers reduce the initial flux of primary hadrons from nucleus-nucleus collisions 
by a factor of 100, and they reduce the low energy particles flow produced in secondary interactions (mainly
low energy electrons). The thick front absorber ($\lambda_{I}\sim10$) is the most critical component and it has been 
designed for minimizing the invariant mass resolution deterioration of the spectrometer due to straggling
and multi-scattering. The spectrometer is also shielded throughout its length by a dense absorber tube
surrounding the beam pipe. A second absorber and four planes of resistive plate detectors are 
used for muon identification and triggering. The front absorber
and the muon filter wall stop muons upto 4 GeV/c. This makes the detection of muons with a momentum 
smaller than 4 GeV/c impossible. 

A total of ten tracking chambers grouped in five stations define the muon trajectories. The tracking stations 
are multi-wire proportional chambers with segmented cathode plane (cathode pad chambers, CPC). 
Each station is made of two chamber planes and each chamber has two CPCs which are both 
read via front-end electronics, in order to have a two-dimensional hit information. 
The detection planes are quadrant structure for stations 1 and 2 and
slat structure for stations 3 to 5. CPCs have been chosen because they can 
be equipped with high granularity read-out and reach the required resolution.
The tracking chambers~\cite{tdr} must satisfy two main requirements: to have a spatial resolution of
about 100 $\mu$m needed for bending ($zy$) plane of the muon trajectory 
to achieve the required mass resolution and to have the capability
to operate, with a good efficiency, in high particle multiplicity environment.  
Two tracking stations are located in front of the dipole magnet, 
another one is located inside, and two more stations sit behind the magnet. 

The front-end electronics~\cite{tdr} is based for all the tracking stations on a 64 channels board 
(MANU) including four charge amplifier chips (MANAS), ADCs and a readout chip allowing zero suppression (MARC). About
17,000 MANU boards are necessary to read the million channels. The data are transferred
from the MANU to the Concentrator ReadOut Cluster Unit System (CROCUS). 
Twenty CROCUS are needed for the five stations. The main tasks
of the CROCUS are to concentrate the data from a half detection plane and send it to the Data Acquisition (DAQ),
to perform the calibration of the MANU and to dispatch the trigger signal from the central
trigger processor. 

The trigger system of the muon spectrometer consists in four planes of Resistive Plate Chamber (RPC) 
detectors operated in streamer mode. A total of 
21,000 Resistive plate chamber front-end channels and fast-decision electronics, 
covering an area of 140 m$^2$ are arranged in two stations one metre apart from each other and placed
behind the muon filter. It is designed to identify the (muon) tracks, in a large background environment, 
to provide a fast trigger signal. Most of the muons emitted in the decays of the quarkonia resonances 
have a transverse momentum $p_{T}$ larger than that of the background muons from pion 
and kaon decays. A cut on the transverse momentum of the muons is done by the trigger 
to reject the background and to select the interesting events,
containing at least a pair of opposite-sign muons with a high $p_{T}$.
From the front-end electronics the signals are send to the trigger electronics, based on 
programmable circuits, working at 40 MHz. 
According to simulations, the total dimuon trigger rate is expected not to exceed 1 kHz for all colliding systems. 
Such a rate complies with DAQ and HLT (High Level Trigger) requirements.

\begin{figure}
\begin{center}
\includegraphics[width=34pc]{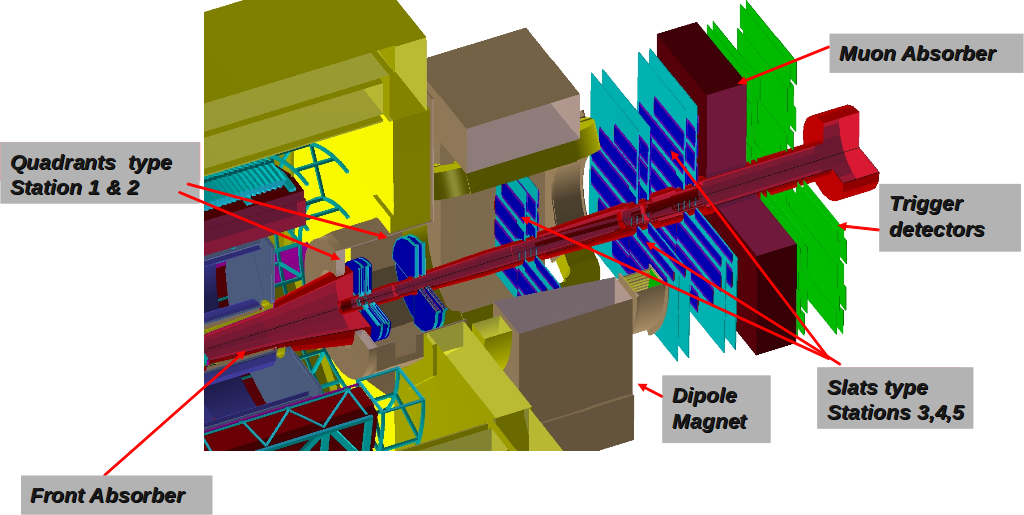}
\caption{(Color Online) The layout of the ALICE Muon Spectrometer. 
The tracking system consists of stations 1 to 5.}
\end{center}
\end{figure}

All the detectors were installed and commissioning started at the beginning of 2007 in
the cavern with the testing of all the functionalities: HV, LV, cooling, gas, electronic noise,
pedestals and stability in time. Next, the detectors ready for the first ALICE cosmic runs were
integrated in the global DAQ with a trigger given by scintillators placed horizontally on the
top of the L3 magnet. The first cosmic tracks in the muon spectrometer were observed in the tracking stations 
1 and 2 and the trigger RPCs during February 2008 when the muon trigger was added to the general
ALICE trigger. During 2010, ALICE collected data in pp collisions at $\sqrt{s}$ = 7 TeV, where 
all the five stations of the Muon Spectrometer participated with a tracking efficiency of $\sim$ 94$\%$. 
Achieving a stable configuration with DAQ and triggers, the Muon Spectrometer recorded their crucial 
magnetic-field off run in May 2010, for the alignment of the detector-elements. This has further 
enhanced the quarkonia measurements by a large scale which is now close to what is expected from simulations.

The data sample for the Muon Spectrometer comprises of Minimum Bias (MB) events and events with at least one 
triggered muon. The main trigger selection was a MB one which is defined by the inclusion of at least a signal in either of
the two scintillator hodoscopes, the V0 detectors, positioned in the forward and backward regions of the ALICE experiment,
or in the pixel barrel detector. This trigger selects events with at least 
one charged particle in eight units of pseudorapidity and is sensitive to about 95$\%$ 
of the pp inelastic cross section~\cite{jpsi}. A second trigger selection ($\mu$-MB) 
requests, in addition to the minimum-bias condition, at least one particle which has been detected in 
the Muon Spectrometer, having a $p_{T}$ larger than the trigger threshold (0.5 GeV/c). The results that we present are 
obtained from data recorded during the initial 2-3 months of pp collisions at $\sqrt{s}$=7 TeV with an 
average luminosity kept at (0.6-1.2)$\cdot10^{29}\rm{cm^{\rm{-2}}s^{\rm{-1}}}$ in order to control the collision pile-up rate 
in the ALICE experiment. An Event display showing tracks in the Muon Spectrometer in 
pp collisions is shown in Fig. 2(left).

\begin{figure}
\includegraphics[width=18pc,height=13pc]{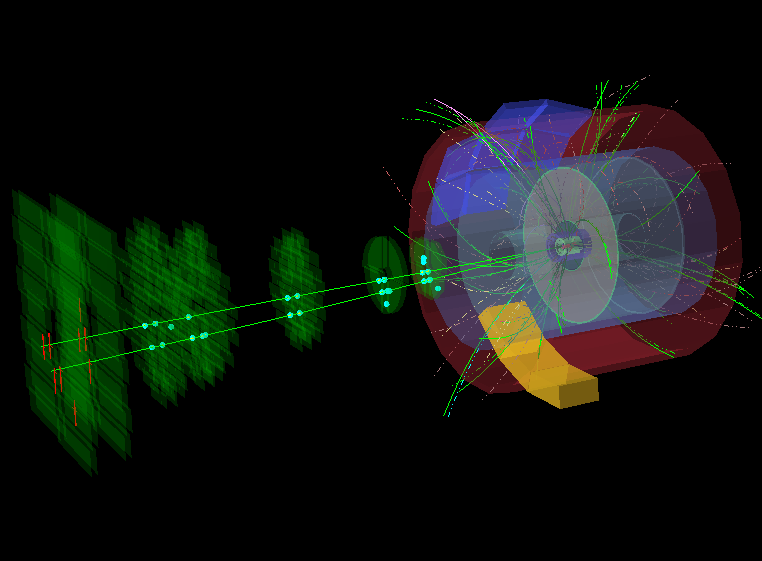}
\vspace{1.0cm}\hspace{0.5cm}
\includegraphics[width=21pc,height=16pc]{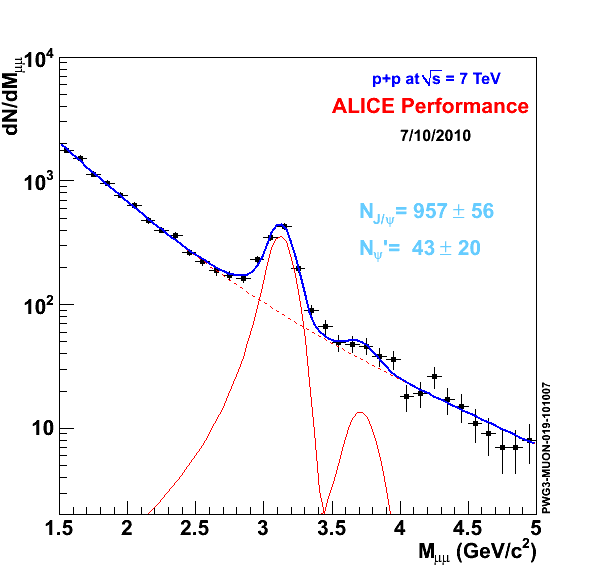}
\caption{(Color Online)Left Panel: ALICE Event display showing muon tracks 
in pp collisions at $\sqrt{s}$=7 TeV. Right Panel: Invariant opposite sign dimuon mass spectrum 
of a sub-set of the data taken in pp collisions in 2010.}
\end{figure}

\section{Quarkonium production in forward rapidities}
\label{}

        For the muon analysis, the results presented here are based on a sample corresponding to an integrated luminosity 
of L = 13.6 nb$^{-1}$~\cite{jpsi}. In this analysis at least one of the two muon candidates is 
required to match the corresponding hits in the trigger chambers. 
A cut on the reconstructed position at the end of the front absorber is applied to each track.
Muons emitted at small angle and which have crossed a significant fraction of the thick beam shield surrounding the beam
pipe, are rejected. Events very close to the edge of the spectrometer acceptance are removed and the cut 
-4.0 $<$ $y$ $<$ -2.5 on the pair rapidity has been applied. The invariant mass spectrum corresponding to a sub-set of our data
is shown in Fig. 2 (right). The J/$\psi$ peak is clearly noticed in the spectrum. The number of
signal events ($N_{J/\psi}$ ) has been extracted by fitting the mass range 1.5 $<$ $m_{\mu\mu}$ $<$ 5 GeV/$c^2$ . The J/$\psi$ 
and $\psi$(2S ) line shapes are fitted with Crystal Ball functions and the underlying 
continuum is described using the sum of two  exponentials. The functions which depict the resonance shapes have 
been obtained by fitting the expected mass distribution of a pure Monte Carlo (MC) signal
sample. This is based on a realistic description of the detectors’ response. Accounting for the small uncertainties in the
MC description of the set-up, the position of the J/$\psi$ mass pole along with the width of the Crystal Ball function are kept as free parameters in the invariant mass fit. Due to the small statistics, the $\psi$(2S ) parameters have been clubbed 
to the J/$\psi$ ones. As shown in Fig. 2 (right) the fit shows a satisfactory $\chi^{2}/ndf$ = 1.14. 
The pole position of the J/$\psi$ is $m_{J/\psi}$ =3.118 $\pm$ 0.005 GeV/$c^2$. 
The extracted width of the Crystal Ball function is $\sigma_{J/\psi}$ =94 $\pm$ 8 MeV/c$^2$ . The total number of 
J/$\psi$ signal events is calculated by integrating the Crystal Ball function over the full mass. 
In this analysis $N_{J/\psi}$ is 1924 $\pm$ 77(stat.).

A MC procedure is used to calculate the acceptance of the apparatus along with the reconstruction and triggering 
efficiencies. These are required to calculate the  J/$\psi$ where the number of signal events need to be corrected 
by these above mentioned factors. The $p_{T}$ distribution used in this MC procedure is extrapolated from CDF measurements 
while the rapidity distribution is taken from  Color Evaporation Model (CEM) calculations~\cite{Vogt:2010aa}. 
The J/$\psi$ production has also been assumed to be unpolarized. To compute the production cross section value, the ratio 
$N_{J/\psi}^{cor}$ = $N_{J/\psi}$ /( $A \times \epsilon$) needs to be normalized to the integrated
luminosity, or to the measured cross section for a chosen reference process. 
For this analysis, the chosen reference process is the MB condition.

In the muon channel the cross-section is calculated as,
\begin{equation}                                                               
\sigma_{J/\psi}(-4.0 < y < -2.5) = \frac{N^{corr}_{J/\psi}(-4.0 < y < -2.5)}{BR(J/\psi\rightarrow \mu^{+}\mu^{-})}\times \frac{\sigma_{MB}}{N_{MB}} \times R 
\end{equation}                                                               
    where BR($J/\psi\rightarrow \mu^{+}\mu^{-}$)=5.93$\%$~\cite{pdg}, $N_{MB}$ is the number of minimum bias collisions, 
corrected for the probability of having multiple interactions in a single bunch crossing, and $\sigma_{MB}$ is the 
measured cross section for such events. In the muon channel, the J/$\psi$  signal has been collected using the $\mu-MB$ 
trigger condition. Therefore, Eq. 1 has an additional multiplicative factor R that links the occurrence of a 
reference process in the $\mu-MB$ and MB event samples. We have chosen as a reference process the 
yield ($N_{\mu}$) of single muons, detected in the region $-4.0 < y < -2.5$, and having $p_{T}>$ 1 GeV/c. 
The R-factor then becomes R = $N_{\mu}^{MB}/N_{\mu}^{\mu-MB}$. The $\sigma_{MB}$ value has been obtained 
relative to the cross-section  $\sigma_{\rm{V_{0}AND}}$ measured in a Van der Meer scan~\cite{vander}, of 
the coincidence between signals in the two VZERO detectors. The error on luminosity measurement 
is dominated by a 10$\%$ systematic error on the determination of $\sigma_{MB}$.
The systematic error on the inclusive J/$\psi$ cross section measurements has been obtained considering the uncertainty on the
signal extraction procedure, the dependence of the acceptance calculation on the input $p_{T}$  and rapidity distributions 
of the J/$\psi$ decay to lepton pairs, the error on the measured luminosity 
and the acceptance dependence on the unknown J/$\psi$ polarization. 
The uncertainties associated with the J/$\psi$ polarization have been calculated in the two cases of 
fully transverse ($\lambda = 1$) or longitudinal ($\lambda = -1$) polarization, in the Collins-Soper (CS) and 
Helicity (HE) reference frames. Quadratically combining the errors from the sources described above, 
except polarization, we obtain a 13$\%$ systematic error for the dimuon channel.

\begin{figure}
\includegraphics[height=16pc,width=20pc]{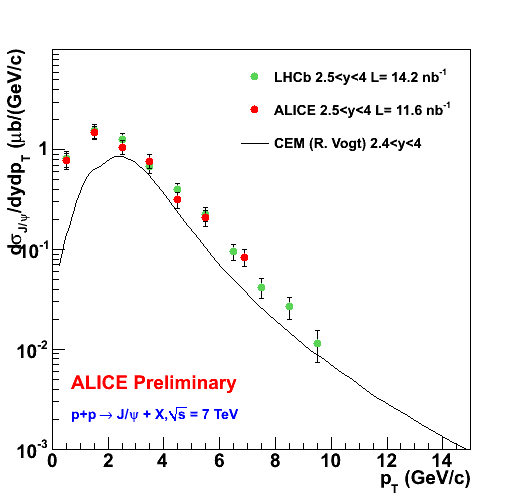}
\includegraphics[height=15pc,width=19pc]{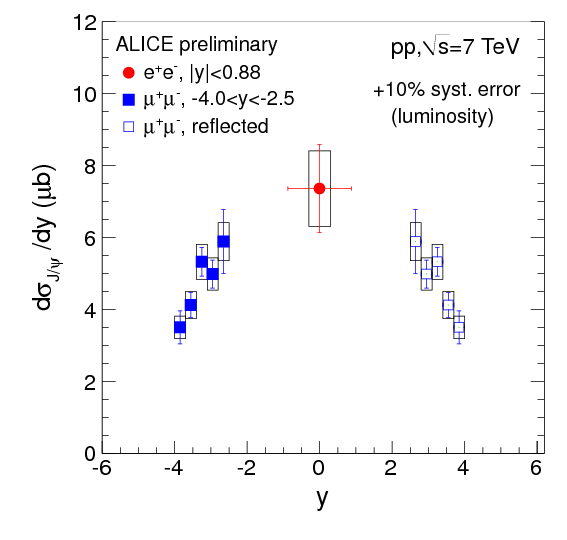}
\caption{(Color Online) Left Panel: Inclusive $J/\psi$ production cross-section as a function 
of $p_{T}$ measured by the ALICE Muon Spectrometer in pp collisions at $\sqrt{s}$=7 TeV. The inclusive cross-section 
measured by the LHCb experiment in the same rapidity range~\cite{lhcb} is also placed for comparison. 
Right Panel: Inclusive $J/\psi$ production cross-section as a function 
of rapidity in pp collisions at $\sqrt{s}$=7 TeV.}
\end{figure}

The resulting inclusive J/$\psi$ production cross sections in pp collisions at $\sqrt{s}$=7 TeV is:
d$\sigma_{J/\psi(-4<y<-2.5)}=7.25\pm0.29(stat.)\pm0.98(syst.)+0.87 (\lambda_{HE} = 1) - 1.50 (\lambda_{HE} = -1)~\mu$b~\cite{jpsi}. 
The quoted systematic errors are related to the unknown polarization values relative to the HE reference frame. 
The cross section value for the muon analysis can be directly compared to the preliminary 
one quoted by the LHCb experiment 
and referring to the same rapidity range (7.65 $\pm$ 0.19 (stat.) $\pm$ 1.10 (syst.) + 0.87 - 1.27 (syst. pol.)
$\mu$b)~\cite{lhcb}. 
The available statistics has allowed the study of the $p_{T}$ and $y$ differential distributions for the di-muon channel. 
In particular, d$\sigma_{J/\psi}/dp_{T}$ has been studied in seven bins between 0 and 8 GeV/c, and 
d$\sigma_{J/\psi}/dy$ in five bins between 2.5 and 4. 

The event sample used for this differential cross-section determination corresponds to L = 11.6 nb$^{-1}$. 
The same fitting technique is used for the
integrated invariant mass spectra to extract the $J/\psi$ signal yield in both the $y$ and $p_{T}$ bins. 
The acceptance $\times$ efficiency has been calculated differentially in 
rapidity and $p_{T}$. For J/$\psi$ production, the A $\times$ efficiency (or $\epsilon$) coverage, 
extends down to zero $p_{T}$, and the values vary by less than a factor 1.6 in the analysed $p_{T}$ 
range. $A \times \epsilon$ has a stronger rapidity dependence, but its values are larger than 10$\%$ 
everywhere. The differential cross sections
are calculated with the same approach used for the integrated cross section, normalizing $N^{cor}_{J/\psi} (y)$ 
and $N^{cor}_{J/\psi}(p_{T})$ to the integrated luminosity and are affected by the same systematic error sources. 
In Fig. 3(left) the results on d$\sigma_{J/\psi}/dydp_{T}$ are presented. 
A comparison with a Color Evaporation Model (CEM) calculation for prompt J/$\psi$ is also shown. 
The prediction underestimates the experimental results, especially at low $p_{T}$, but a more
meaningful comparison would require the subtraction of the contribution from b-decays from the data.
The d$\sigma_{J/\psi}/dy$ results in the electron and dimuon channel are shown in Fig. 3(right). 
Excluding the effect of polarization the systematic error of the dielectron measurement is 16$\%$.
The values obtained in the forward region have also been reflected with respect to y = 0.

\begin{figure}
\begin{center}
\includegraphics[height=20pc]{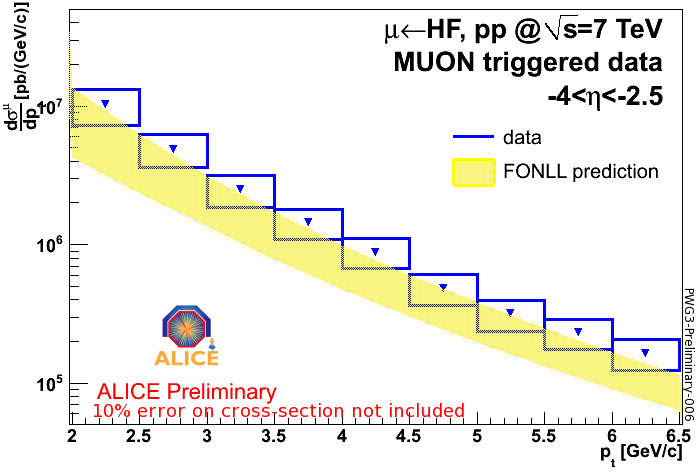}
\caption{(Color Online)The differential transverse momentum cross-section for muons from 
heavy-flavor decays in the rapidity domain of -4.0 $<$ $\eta$ $< -2.5$. The statistical error is small (hidden in the markers) 
and the systematic errors (boxes) do not include an additional $10\%$ error on the minimum-bias pp cross-section.
The comparison with the FONLL prediction~\cite{pvt} is also shown.}
\end{center}
\end{figure}

\section{Cross-section of forward single muons from heavy-flavor decays}
\label{}

The results presented here are based on a sample corresponding to an integrated luminosity of L = 3.49 nb$^{-1}$~\cite{single}. 
The extraction of the heavy-flavor contribution from 
the single muon spectra requires the subtraction of background sources which constitute of three main contributions. They 
are as follows: a) muons which are from the decay-in-flight of light hadrons produced at the pp interaction point(decay muons); b) muons produced after hadronic interactions with absorber materials (secondary muons); c) punch-through hadrons.
The last contribution is rejected by requiring the matching of the reconstructed tracks with trigger system tracklets. 
As the parent particles have lower mass, the background muons have a softer $p_{T}$ 
spectrum compared to the heavy-flavor muons. The background particles dominate the low-$p_{T}$ region. 
The analysis focuses on the region 2 $<$
$p_{T}$ $<$ 6.5 GeV/c where the upper limit is influenced by the $p_{T}$  resolution of spectrometer. An initial, partial,
alignment used in this analysis is done with a sample of tracks collected without the magnetic field in the dipole. 
The estimated resolution for the coordinate perpendicular to the magnetic field, 
which is measured with precision, ranges from 300 $\mu$m to 
900 $\mu$m, depending on the detection element~\cite{single}. 
The contribution of secondary muons is small
(about 3$\%$) in this transverse momentum range. 
The main source of background in this region comprises of decay muons (about 25$\%$), which have been
subtracted using simulations. The transverse momentum distribution was produced with the Perugia-0 tune
of PYTHIA~\cite{pythia}. This distribution is normalized in such a manner that 
the resulting fraction of decay-muons in the $p_{T}$ range where this
contribution is dominant ($0.5 <$ $p_{T}$ $< 1$ GeV/c) is similar to the one provided by the simulation. The systematic error
introduced by this procedure was estimated as 30$\%$ to 20$\%$ (from low to high $p_{T}$) by comparing the results obtained
with other tunings of PYTHIA and by varying by 100$\%$ the secondary contribution from the event generator.
    After background subtraction, the muon $p_{T}$ spectrum is corrected for efficiency ($\sim$80$\%$) and
normalized to a cross section using the minimum-bias pp cross section measured, with a Van der Meer scan~\cite{vander}. 
The charm and beauty decay muon cross section d$\sigma$/d$p_{T}$ 
in the range 2 $<$ $p_{T}$ $<$ 6.5 GeV/c and $-4 <$ $\eta$ $< -2.5$ is presented in Fig. 4. 
The corresponding FONLL pQCD calculation~\cite{pvt} agrees with our data within uncertainties. 
The next step in this analysis is to extend the $p_{T}$ reach up to about 20 GeV/c,
using data collected since summer 2010 that have been reconstructed with improved alignment corrections for the
tracking chambers. 
A dominance of the beauty decays in the single-muon cross-section is expected in the range of 
$10 <$ $p_{T}$ $< 20$ GeV/c; thus, this measurement will lead towards the reference for the study of bottom quark quenching at
forward rapidity in heavy-ion(Pb-Pb) collisions.

\begin{figure}
\begin{center}
\includegraphics[height=22pc]{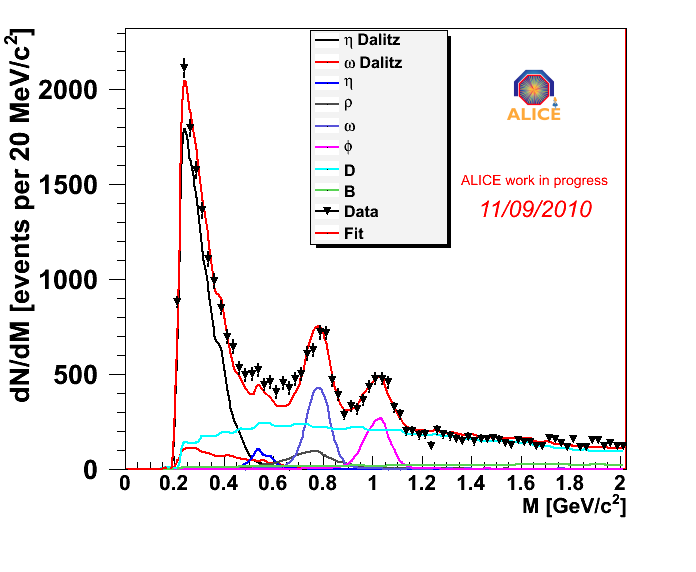}
\caption{(Color Online) Low mass vector mesons with the corresponding Monte-Carlo fits in the di-muon 
opposite sign mass spectrum. The errors are purely statistical.}
\end{center}
\end{figure}

\section{Low-mass Vector Mesons}
\label{}

For the study of low mass vector mesons the tracks reconstructed in the Muon Spectrometer 
are required to match with the tracklets in the muon trigger. 
A cut on the muon rapidity $-4 < y < -2.5$ is also applied, to exclude the region at the border of the spectrometer acceptance.
For a given event sample the opposite sign muon track pairs $N^{+-}$, and like sign 
muon track pairs ($N^{++}$  and  $N^{--}$) are reconstructed. 
Then, for that sample,the mixed event spectrum (which gives only the shape, and actually contains a number of pairs which is in
principle arbitrary and anyway much bigger than the actual background) is rescaled such that 
its integral equals $2\sqrt{(N^{++}\cdot N^{--})}$, where $N^{++}$ and $N^{--}$ are the numbers of 
muon pairs in the real data with both positive and negative muons respectively. 
To normalize the mass spectrum, since the expected contribution of correlated like-sign pairs is
negligible, it is assumed that the like-sign pairs are uncorrelated. This is an analysis in progress and the 
statistics is sufficient to study the ratio of $\phi/(\rho + \omega)$ as a function of $p_{T}$ along 
with the $\phi$ yields with $p_{T}$. The invariant mass spectrum for low mass vector mesons is shown in Fig. 5. 
The $\phi$ peak and combined peak of $\rho + \omega$ is seen after background subtraction. 
Yields and the transverse momentum spectra of $\phi$ and 
$\rho / \omega$ are determined by comparing the subtracted spectrum to a Monte-Carlo simulation of all the processes 
that contribute.

\begin{figure}
\includegraphics[width=19pc,height=14pc]{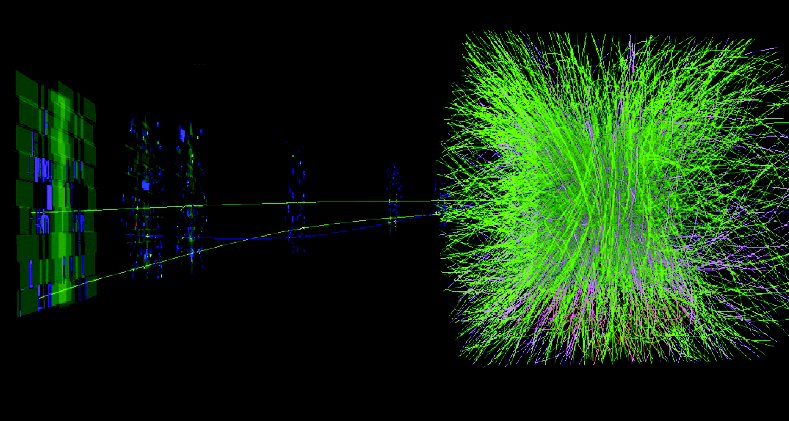}
\vspace{1.0cm}\hspace{0.5cm}
\includegraphics[width=18pc,height=14pc]{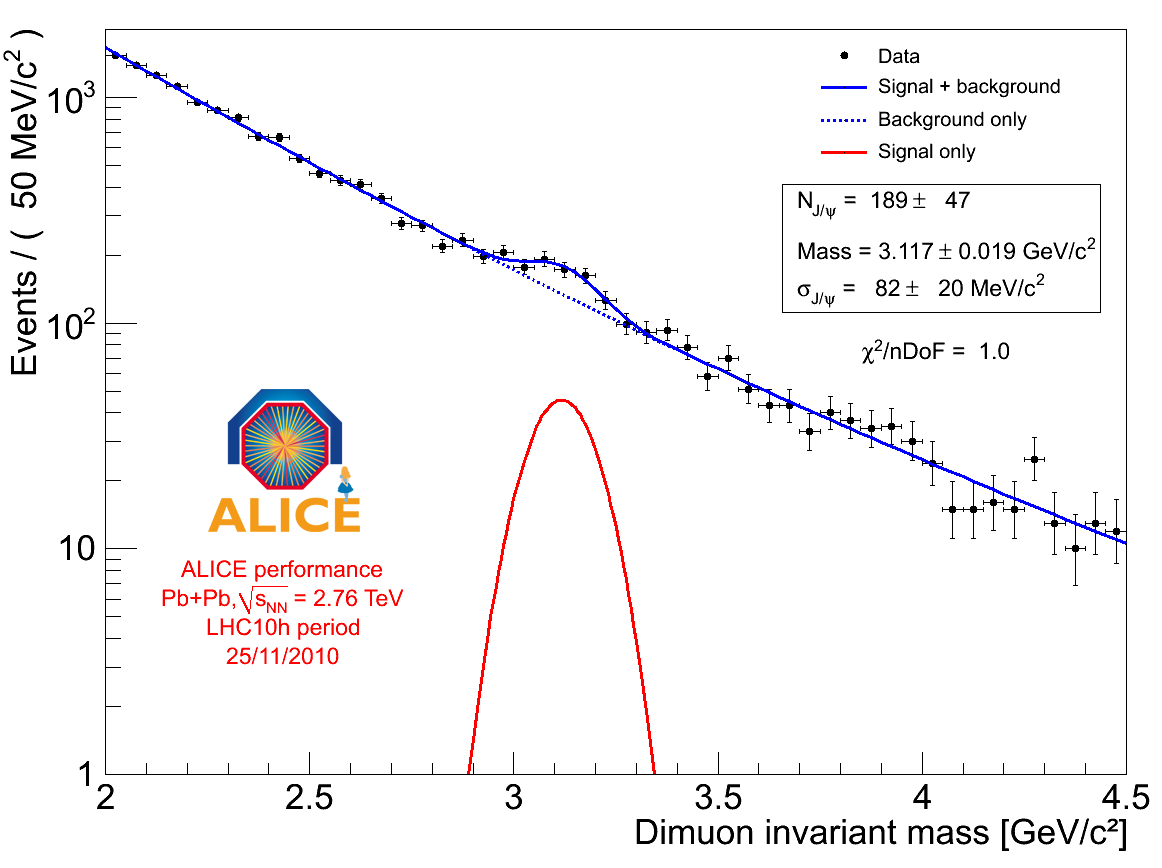}
\caption{(Color Online) Left Panel: ALICE Event display showing muon tracks 
in Pb-Pb collisions at $\sqrt{s_{\rm NN}}$= 2.76 TeV. Right Panel: Invariant opposite sign dimuon mass spectrum 
of a fraction of the data-set taken in 2010 Pb-Pb collisions.}
\end{figure}

\section{Summary and Outlook}
\label{}

The first heavy-flavor production measurements performed by the ALICE Muon Spectrometer in pp collisions at 
$\sqrt{s}$=7 TeV at LHC is presented. The inclusive J/$\psi$ cross-section measured is  
d$\sigma_{J/\psi(-4<y<-2.5)}=7.25\pm0.29(stat.)\pm0.98(syst.)+0.87 (\lambda_{HE} = 1) - 1.50 (\lambda_{HE} = -1)~\mu$b.
The transverse momentum dependence has also been measured in 
the range of $0 < p_{T} < 8$ GeV/c at the forward rapidities. The combined d$\sigma_{J/\psi}/dy$ results of 
di-muon and di-electron channel measured by ALICE show its strength in terms of the broad acceptance and 
sensitivity down to zero $p_{T}$ at all rapidities. The production cross-section of heavy-flavor decay muons 
at forward rapidity has been measured as a function of $p_{T}$ in the range of 2-6.5 GeV/c. Perturbative 
QCD calculations (FONLL) agree with our measurements within uncertainties.

The new era in search of utilizing these heavy quark probes to understand the strongly interacting matter under extreme 
conditions opened from November 2010, when LHC produced the first Pb-Pb collisions in centre-of-mass energy per 
nucleon pair of $\sqrt{s_{\rm NN}}$= 2.76 TeV~\cite{Aamodt:2010pb}. An event display showing tracks in the Muon Spectrometer in 
Pb-Pb collisions in shown in Fig. 6(left). The invariant mass spectrum corresponding to a fraction of our heavy-ion 
data is shown in Fig. 6(right).
The full analysis is in progress and the results depicting an unknown energy regime with an   
increase of more than an order of magnitude over the highest energy 
nuclear collisions previously obtained in the laboratory, will reveal new understanding.


\begin{thebibliography}{00}

\bibitem{Fabjan:2011jb}
  C.~Fabjan and J.~Schukraft,
  arXiv:1101.1257 [physics.ins-det]; K. Aamodt et al. [ALICE Collaboration], JINST 3 (2008) S08002; 
F. Carminati et al. [ALICE Collaboration], Physics Performance Report Vol. I, CERN/LHCC 2003-049 and J. Phys. G30 1517 (2003); 
B. Alessandro et al. [ALICE Collaboration], Physics Performance Report Vol. II, CERN/LHCC 2005-030 and J. Phys. G32 1295 (2006).

\bibitem{Shuryak:1978ij}
  E.~V.~Shuryak,
  Phys.\ Lett.\  B {\bf 78}, 150 (1978)
  [Sov.\ J.\ Nucl.\ Phys.\  {\bf 28}, 408 (1978)]
  [Yad.\ Fiz.\  {\bf 28}, 796 (1978)].

\bibitem{Braaten:1991we}
  E.~Braaten and M.~H.~Thoma,
  Phys.\ Rev.\  D {\bf 44}, 2625 (1991); H.~van Hees, V.~Greco and R.~Rapp,
  Phys.\ Rev.\  C {\bf 73}, 034913 (2006)
  [arXiv:nucl-th/0508055]; D.~Das et al. [STAR Collaboration],
  Eur.\ Phys.\ J.\  C {\bf 62}, 95 (2009); 
  A. Adare et al. [PHENIX Coll.], Phys. Rev. Lett. 97 (2006) 252002.



\bibitem{Matsui:1986dk}
  J.~J.~Aubert {\it et al.}  [E598 Collaboration],
  Phys.\ Rev.\ Lett.\  {\bf 33}, 1404 (1974); T.~Matsui and H.~Satz,
  Phys.\ Lett.\  B {\bf 178}, 416 (1986).



\bibitem{physreptvogt}
  A.~D.~Frawley, T.~Ullrich and R.~Vogt,
  Phys.\ Rept.\  {\bf 462}, 125 (2008)
  [arXiv:0806.1013 [nucl-ex]];
  R.~Vogt,
  Phys.\ Rept.\  {\bf 310}, 197 (1999).


\bibitem{tdr}The Dimuon Forward Spectrometer Technical Design Report. CERN/LHCC 99-22; 
The Dimuon Forward Spectrometer TDR Addendum. CERN/LHCC 2000-046; B.~Espagnon  [ALICE Collaboration],
  J.\ Phys.\ G {\bf 35}, 104145 (2008).

 
\bibitem{tevatron}

M. Cacciari et al., JHEP 0407 (2007) 033.
N. Brambilla et al. (Quarkonium Working Group), arXiv:1010.5827v1[hep-ph], accepted in EPJC.
 
\bibitem{jpsi}
 A. Dainese et al. [ALICE Collaboration], Hard Probes-2010[arXiv:1012.4036[hep-ex]];
 J. Schukraft et al. [ALICE Collaboration], these proceedings;
 R. Arnaldi et al. [ALICE Collaboration], Hard Probes-2010 proceedings; S. Pal et al. [ALICE Collaboration], these proceedings.




\bibitem{Vogt:2010aa}
  R.~Vogt,
  Phys.\ Rev.\  C {\bf 81}, 044903 (2010) and private communication; 
  D. Stocco et al., ALICE Internal Note ALICE-INT-2006-029.


\bibitem{pdg}
K. Nakamura et al. (Particle Data Group), J. Phys. G 37, 075021 (2010) 

\bibitem{vander}
S. Van der Meer, ISR-PO/68-31, KEK68-64.


\bibitem{lhcb}
G. Passaleva et al. [LHCb Collaboration], presented at ICHEP 2010, Paris, July 2010.


\bibitem{single}
 J. C. Castellanos et al. [ALICE Collaboration], ICHEP 2010 proceedings; 
 D. Stocco et al. [ALICE Collaboration], Hard Probes-2010 proceedings; 

     
\bibitem{pythia}
 P. Z. Skands, Phys. Rev. D82 (2010) 074018.


\bibitem{pvt}
 M. Cacciari et al., private communication.
     


\bibitem{Aamodt:2010pb}
  K.~Aamodt {\it et al.}  [ALICE Collaboration],
  Phys.\ Rev.\ Lett.\  {\bf 105}, 252302(2010)],
  arXiv:1011.3916 [nucl-ex].


\end{thebibliography}
\end{document}